\documentclass[pra,prl,twocolumn,superscriptaddress,floatfix,showpacs,10pt]{revtex4}
\usepackage{graphicx}
\usepackage{slashbox}
\usepackage{amssymb}
\usepackage{verbatim}

\begin{document}

\title{Characterizing the complete hierarchy of correlations
in an $n$-party system}

\author{D.L. Zhou}
\affiliation{Institute of Physics, Chinese Academy of Sciences,
Beijing 100080, China}

\author{L. You}
\affiliation{School of Physics, Georgia Institute of Technology,
Atlanta, Georgia 30332, USA}

\begin{abstract}
A characterization of the complete correlation structure
in an $n$-party system is proposed in terms of
a series of $(k,n)$ threshold classical
secret sharing protocols ($2\le k\le n$).
The total correlation is shown to be the sum of
independent correlations of $2$-, $3$-,$\cdots$, $n$-parties.
Our result unifies several earlier scattered works,
and shines new light at the important topic of multi-party
quantum entanglement. As an application, we explicitly construct the
hierarchy of correlations in an $n$-qubit graph state.
%We also provide a comprehensive understanding of correlation
%properties within all three-qubit pure states.
\end{abstract}

 \maketitle

Despite their wide usage, correlations, especially
more than two-party correlations in a multiparty
system remain to be fully understood.
Two-party correlation is relatively well understood.
It is typically measured by the two-party
mutual entropy \cite{henderson}, which gives the
remarkable result that the two-party correlation
of a Bell state is twice
that of the maximally correlated classical two-qubit state.
Groisman \textit{et. al.}, provided the first
operational interpretation for two-party
correlation \cite{Groisman} based on
the idea of Landauer---the amount of information
equals the amount of work required for its erasure \cite{Landauer}.
More recently, Schumacher and Westmoreland
published a direct proof equating the
two-party mutual entropy to
the maximal amount of information that one party can send to the
other in a one-time pad cryptography \cite{Schumcher}.

For a Bell state, the cryptographic scheme of
Schumacher and Westmoreland is simply
the superdense coding protocol \cite{Bennett}
where two bits of classical information are
communicated by transmitting one qubit.
Different roles, the sender and
the receiver, respectively, are assumed by the two parties
in this protocol. The correlation
between the two parties, however, is symmetric,
\textit{i.e.}, it makes no sense to phrase that
the correlation is from one party (sender) to the other (receiver).
The same has to hold for multi-parties, {\it i.e.},
any operational definition for the degree of multi-party
correlation has to be symmetric with respect to all parties.

The above discussion of correlation echoes
multiparty secret sharing (SS) schemes,
both are symmetric with respect to all parties.
In 1979, Blakely \cite{Blakely}
and Shamir \cite{Shamir} addressed the issue of a
$(k,n)$ threshold protocol for sharing a secret
that can be recovered by $k$ or more parties,
but not by less than $k$ parties.
Quantum SS was first discussed by M.
Hillery \textit{et. al.}, and was associated
with establishing classical or quantum secret keys
among the multi-parties \cite{Hillery}.
Cleve {\it et. al.}, proposed an improved $(k,n)$ threshold
quantum SS protocol \cite{Cleve}, which allowed
an unknown quantum state to be shared
in a multi-party system and made
connections to quantum error correction codes.
Terhal \textit{et. al.}, presented a scheme
for hiding a classical bit into a collection
of Bell states (between two parties),
allowing for all types of classical communication,
but not two-party quantum communication \cite{Terhal}.
Eggeling and Werner generalized the
protocol of Terhal \textit{et. al.}, to
multiparties \cite{Eggeling} and pointed out
that quantum entanglement is not required.

The intimate connection between correlations
and SS protocols has led to the present work,
where we propose a
complete characterization of the
hierarchy of correlations in an $n$-party system
with a series of $(k,n)$ ($2\le k\le n$) threshold
classical SS protocols.
This Letter is organized as follows.
We start by revisiting the simplest possible
case of a two-party state, fully analyzing
its correlation in terms of its SS capacity.
The definition for the total correlation in
an $n$-party state then naturally arises.
Using examples of three-party states,
the total correlation is shown to be composed
of independent two- and three-party correlations,
which paves the way for a proper generalization
to the complete correlation hierarchy of $n$-party states.
Before summarizing our result, we provide an
explicit construction of the complete
correlation structures for all graph states.

The maximally correlated classical two-qubit state
\begin{eqnarray}
\rho^{(12)}_{c}=\frac{1}{2}\left(|00\rangle_{12}\
_{12}\!\langle 00|+|11\rangle_{12}\
_{12}\!\langle 11| \right),
\label{2ce}
\end{eqnarray}
gives rise to completely mixed
reduced density matrices for both parties
$j=1,2$, \textit{i.e.}, $\rho^{(j)}=I_j/2$.
This state gives a random outcome ``$+1$" or ``$-1$"
with equal probabilities when each
qubit is measured independently with
the Pauli matrix $Z_j$.
The results from the two qubits, however,
reveal the inherent correlation because
$Z_1 Z_2\equiv1$ for the state (\ref{2ce}).

The above example shows that in an approximate sense,
correlation specifies the definiteness of
a composite state with uncertainties for its parts.
To provide a more precise characterization,
we introduce an alternative picture
for the state (\ref{2ce}) by
defining two logical qubits ${\tilde{1}}$ and ${\tilde{2}}$
with $Z_{\tilde{1}}=Z_1 Z_2$,
$X_{\tilde{1}}=X_1$, $Z_{\tilde{2}}= X_1
X_2$, and $X_{\tilde{2}}=Z_2$.
This gives rise to a transparent form
$\rho^{(12)}_{c}=|0\rangle_{\tilde{1}}\
_{\tilde{1}}\langle 0|\otimes I_{\tilde{2}}/2$,
clearly revealing the presence of one bit of
correlation encoded in the first logical qubit ${\tilde{1}}$.
%From this perspective, correlation is naturally defined
%as the difference in uncertainties for the local parties
%of the whole state.
We now relate this correlation
measure to a classical SS protocol,
where a secret is encoded
by a unitary transformation that leaves the state
invariant for all local parties.
The capacity for SS by a state
is then defined as the maximal
number of secrets encodable
or the maximal number of unitary transformations
that are distinguishable in a single measurement to this state.
The degree of correlation
is then measured by the capacity for SS.
For the state (\ref{2ce})
$\rho_c^{(12)}$, $1$ bit of classical information $c\in
\{0,1\}$ can be encoded into a unitary transformation
$X_{\tilde{1}}^c$, and the secret $c$ can be decoded by a
measurement with $Z_{\tilde{1}}$.

We next consider the Bell state%
\begin{eqnarray}
\left\vert B\right\rangle_{12} =\frac{1}{\sqrt{2}}\left( \left\vert
00\right\rangle _{12}+\left\vert 11\right\rangle _{12}\right),
\label{2eb}
\end{eqnarray}
with the same reduced matrix $\rho^{(j)}=I_j/2$
as for the state (\ref{2ce}).
In terms of the aforementioned two logic qubits, we find
$\left\vert B\right\rangle_{12}=|0\rangle_{\tilde{1}}\otimes
|0\rangle_{\tilde{2}}$, {\it i.e.}, a state capable of encoding
two secret bits $c_1$ and $c_2$ with
unitary transformations $X_{\tilde{1}}^{c_1} $ and
$X_{\tilde{2}}^{c_2}$ that are recoverable from
$Z_{\tilde{1}}$ and $X_{\tilde{2}}$ measurements.
Thus the Bell state (\ref{2eb})
can share $2$ bits of secret, {\it i.e.}, it contains twice
as much correlation as the classical state (\ref{2ce}),
in agreement with the result of Ref. \cite{Groisman}.
Alternatively, the $2$ bits of secret can
be encoded by
$X_{\tilde{1}}^{c_1}=X_1^{c_1}$ and
$(Z_{\tilde{1}}X_{\tilde{2}})^{c_2}=Z_1^{c_2}$.
This latter encoding
involves operations only on a single party,
which gives nothing but the familiar superdense coding protocol \cite{Bennett}.

We aim for a proper two-party correlation measure $C_2(.)$
satisfying the additivity relationship
$C_2(\rho^{(12)}\otimes\sigma^{(12)})=C_2(\rho^{(12)})+C_2(\sigma^{(12)})$
for general states $\rho^{(12)}$
and $\sigma^{(12)}$ shared between the two parties.
This calls for a discussion of
the average degree of correlation for an ensemble
of identical copies of $\rho^{(12)}$ described by
$\rho^{(12)}_{\rm ens}=\prod_{i=1}^{N} \otimes \rho^{(12)}_i$,
whose one-party
reduced density matrix is
$\rho^{(j=1,2)}_{\rm ens}=\prod_{i=1}^{N} \otimes \rho^{(j)}_i$.
According to Schumacher's theorem on noiseless quantum data
compression, the reduced state for party $j$ can be
encoded into $NS(\rho^{(j)})$ qubits with
completely random reduced state in the limit of $N\to\infty$.
The whole state of the ensemble $\rho^{(12)}_{\rm ens}$,
on the other hand, can be encoded into $NS(\rho^{(12)})$
completely random qubits.
The correlation, shared between the two parties,
is exactly the reason for the reduction of
the total number of compressed qubits
from $N(S(\rho^{(1)})+S(\rho^{(2)}))$
to $NS(\rho^{(12)})$.
Therefore, the average correlation for
a general two-party state becomes
\begin{eqnarray}
C_2(\rho^{(12)})=I(\rho^{(12)})\equiv
S(\rho^{(1)})+S(\rho^{(2)})-S(\rho^{(12)}),
\end{eqnarray}
which can be considered as a direct deduction of the main
theorem of \cite{Schumcher}.

Similar argument based on data compression
enables a proper generalization to multiparty states.
We define
\begin{eqnarray}
C_T(\rho^{(12\cdots n)})=\sum_{i=1}^{n}
S(\rho^{(i)})-S(\rho^{(12\cdots n)}),
\label{tc}
\end{eqnarray}
as the total correlation in an $n$-party state $\rho^{(12\cdots
n)}$. In the language of SS, the total correlation
(\ref{tc}) is simply
the capacity for the (2,\,n) threshold classical SS
in an $n$-party state $\rho^{(12\cdots N)}$,
a direct generalization
of the two-party result.
However, the total correlation (\ref{tc}) alone does
not provide sufficient information on the
correlation structure in an $n$-party $(n\ge 3)$ state.
Therefore our further analysis below will concentrate on
characterizing
how the total correlation is distributed among the
$n$ parties.

We now examine the three-qubit maximally
correlated classical state
\begin{eqnarray}
\rho^{(123)} _{c}=\frac{1}{2}\left(|000\rangle_{123}
\ _{123}\langle 000|+|111\rangle_{123}\ _{123}\langle 111|\right),
\label{3ce}
\end{eqnarray}
with $C_T(\rho_c^{(123)})=2$. The two-party correlations
are calculated easily, given by
$C_2(\rho_c^{(12)})=C_2(\rho_c^{(23)})=C_2(\rho_c^{(13)})=1$.
This result leads to an interesting paradox:
the total correlation is less than the apparent
total two-party correlation, \textit{i.e.},
$C_T(\rho_c^{(123)})<C_2(\rho_c^{(12)})+C_2(\rho_c^{(23)})
+C_2(\rho_c^{(13)})$,
a puzzle previously encountered when
three-party mutual entropy was found to be negative
for certain quantum states \cite{Vedral}.
From the view point of our proposed characterization scheme,
the reason for the above paradox
is simple: the three two-party correlations
$C_2(\rho_c^{(12)})$, $C_2(\rho_c^{(23)})$, and $C_2(\rho_c^{(13)})$
are not independent of each other, thus they cannot be simply
added together to give the total two-party correlation.
In fact, the correlation between
the first qubit and the other two qubits is $C_2(\rho_c^{(1(23))})=1$,
causing
$C_2(\rho_c^{(1(23))})<C_2(\rho_c^{(12)})+C_2(\rho_c^{(13)})$,
thus, at most one of the two correlations
$C_2(\rho_c^{(12)})$ and $C_2(\rho_c^{(13)})$ is independent
when the second and the third qubits
are considered as independent parties.
More generally for the state (\ref{3ce}), only two of the
three two-party correlations are independent
when all three qubits are viewed as independent parties.
Any two can be used because the state (\ref{3ce})
is completely symmetric. The
equality $C_T(\rho_c^{(123)})=C_2(\rho_c^{(12)})+C_2(\rho_c^{(13)})$
then excludes the existence of any genuine three-party correlation
in $\rho_c^{(123)}$ and gives rise to the
following simple correlation structure:
the total correlation is $2$ bits,
which is distributed exclusively into any two of the three
two-party correlations of $1$ bit each.

The above correlation structure can be easily understood in the
language of classical SS with the introduction of three
logic qubits $Z_{\tilde{1}}=Z_1 Z_2$,
$X_{\tilde{1}}=X_1$; $Z_{\tilde{2}}= Z_2 Z_3$, $X_{\tilde{2}}=X_2$;
and $Z_{\tilde{3}}= X_1 X_2 X_3$, $X_{\tilde{3}}= Z_3$.
The state (\ref{3ce}) then takes the form
$\rho^{(123)}_{c}= |0\rangle_{\tilde{1}}\; _{\tilde{1}}\langle
0|\otimes|0\rangle_{\tilde{2}}\; _{\tilde{2}}\langle 0|\otimes
I_{\tilde{3}}/2$, capable of encoding
two bits of secret $c_1$ and $c_2$ with $X_1^{c_1} X_2^{c_2}$.
A single measurement with $Z_{\tilde{1}}=Z_1
Z_2$ and $Z_{\tilde{2}}=Z_2 Z_3$ then accomplishes the decoding.
Additionally, we note that the identity of
$Z_{\tilde{1}}Z_{\tilde{2}}=Z_1 Z_3$ allows
for the interchange of the roles for
$Z_{\tilde{1}}$ or $Z_{\tilde{2}}$,
because only two
of the three two-party correlations are independent.

The second three-qubit state we consider
is the three-qubit GHZ state
\begin{eqnarray}
\left\vert G\right\rangle_{123} =\frac{1}{\sqrt{2}}\left( \left\vert
000\right\rangle _{123}+\left\vert 111\right\rangle _{123}\right),
\label{ghz}
\end{eqnarray}
whose total correlation is $C_T(|G\rangle_{123})=3$. Its
two-party correlation structure is the same as that of the
state (\ref{3ce}) because both
share identically the same two-party reduced density
matrices. This then leads to the simple result for the
degree of three-party
correlation of the state (\ref{ghz}) being
$C_T(|G\rangle_{123})-C_T(\rho_c^{(123)})=1$,
a result easily understood again in terms of
SS.
Using the same set of three logic qubits introduced above,
we find $\left\vert
G\right\rangle_{123}=|0\rangle_{\tilde{1}}\otimes
|0\rangle_{\tilde{2}}\otimes |0\rangle_{\tilde{3}}$,
capable of coding three bits of
classical secret $c_1$, $c_2$, and $c_3$
with $X_1^{c_1}X_2^{c_2}Z_3^{c_3}$.
The decoding is achieved analogously
by measurements with
$Z_{j=\tilde{1},\tilde{2},\tilde{3}}$.
Clearly, $Z_{\tilde{3}}$ probes
three-party correlation that cannot be detected
by any two-party measurement.

The total correlation in a two-party
state is simply the two-party correlation.
For a three-party state, however,
the total correlation includes both
three-party correlation and independent
two-party correlation. The calculation
of this independent two-party correlation
is generally a mathematically challenging
task, as evidenced by why the three-party
mutual entropy, defined by
\begin{eqnarray}
I(\rho^{(123)})=C_T(\rho^{(123)})&-&
C_2(\rho^{(12)})\nonumber\\
&-&C_2(\rho^{(13)})-C_2(\rho^{(23)}),
\end{eqnarray}
is not an appropriate measure for three-party correlation.
Expressing it as
$I(\rho^{(123)})=C_2(\rho^{(1(23))})-C_2(\rho^{(12)})-C_2(\rho^{(13)})$,
formally analogous to the two-party mutual entropy,
one might be tempted to consider $I(\rho^{(123)})$
as a reasonable three-party correlation measure.
Yet, this is generally unacceptable because
the two-party
correlations $C_2(\rho^{(12)})$ and $C_2(\rho^{(13)})$
are not always
independent when the two-party correlation
$C_2(\rho^{(1(23))})$ is considered.
%For instance, the state
%$\rho_c^{(123)}$ of (\ref{3ce}) considered before
%displays no three-party correlation, or
%$C_3(\rho_c^{(123)})=0$, although it
%gives $C_2(\rho_{c}^{(1(23))})=1$ and
%$C_2(\rho_c^{(12)})+C_2(\rho_{c}^{(13)})=2$, leading to
%a negative $I(\rho^{(123)})=-1$.

The mathematical difficulty of classifying
independent multiparty correlations can be
resolved with our proposed classification
scheme based on classical SS,
although the actual computation for general
multiparty states may still become completely out of reach.
A three-party state $\rho^{(123)}$ can admit
the $(2,3)$ and $(3,3)$ threshold classical SS
protocols in general.
Our definition (\ref{tc}) for the total
correlation in a three-party state
reduces simply to the capacity of the $(2,3)$
classical secret sharing.
The capacity for the $(3,3)$ threshold
classical secret sharing measures nothing but
the three-party correlation.
The total (independent) two-party correlation
can then be
obtained as equal to the difference between
the total correlation and the three-party correlation.

Formally the definition
for the capacity of the $(k,n)$ threshold classical SS
is based on an ensemble of identical copies of $n$-party
state $\rho^{(12\cdots n)\otimes N}$.
$M_k$ secret bits $\{c_m\} \; (c_m\in
\{0,1\}\; \mathrm{and}\; m=1,2,\cdots,M_k)$ are encoded
with a series of unitary transformations $U(\{c_m\})$,
that leave all reduced density matrices of $(k-1)$-party
invariant, \textit{i.e.}, $\forall
S^k_j$,
\begin{eqnarray}
\mathrm{Tr}_{S^k_j} \left[U(\{c_m\})\rho^{(12\cdots n)\otimes
N}U^{\dagger}(\{c_m\})\right]=\mathrm{Tr}_{S^k_j} \rho^{(12\cdots n)\otimes N},
\end{eqnarray}
where $S^k_j=\{j_\alpha|\; \alpha\in\{1,2,\cdots, n-k+1\},
\; j_\alpha\in\{1,2,\cdots,n\}\}$.
If the secret bits are decoded by
a single measurement,
all coded states
$U(\{c_m\})\rho^{(12\cdots n)\otimes N}U^{\dagger}(\{c_m\})$
are required to be orthogonal to each other.
The capacity of this $(k,n)$
threshold classical SS for the state $\rho^{(12\cdots n)}$
is then given by
\begin{eqnarray}
C^{(k,n)}(\rho^{(12\cdots n)})=\lim_{N\to\infty} \frac {\max M_k} {N}.
\end{eqnarray}
The total correlation of the state $\rho^{(12\cdots n)}$ is
defined as
\begin{eqnarray}
C_T(\rho^{(12\cdots n)})=C^{(2,n)}(\rho^{(12\cdots n)}),
\end{eqnarray}
consistent with our earlier definition (\ref{tc}).
The $k$-party $(2\le k\le n-1)$ correlation is given by
\begin{eqnarray}
C_k(\rho^{(12\cdots n)})=C^{(k,n)}(\rho^{(12\cdots n)})
-C^{(k+1,n)}(\rho^{(12\cdots n)}),
\end{eqnarray}
and the $n$-party correlation is
\begin{eqnarray}
C_n(\rho^{(12\cdots n)})=C^{(n,n)}(\rho^{(12\cdots n)}).
\end{eqnarray}
Our classifying scheme then leads to
\begin{eqnarray}
C_T(\rho^{(12\cdots n)})=\sum_{k=2}^{n}C_k(\rho^{(12\cdots n)}).
\end{eqnarray}

%Three remarks are now in order
%on the correlation structure in a thr-party state:
%First, the total correlation in the state
%$\rho^{(123)}$ equals to the sum of the two-party and the
%three-party correlations; Second, for a general three-party state,
%the calculation of the three-party correlation according to the
%$(3,3)$ threshold secret sharing as discussed above is difficult,
%although the total correlation defined by
%(\ref{tc}) is rather easy to evaluate;
%Third, the above classification scheme
%for the total correlation in a three-party state
%can be directly generalized to the $n$-party case.
%
%For an $n$-party state, our classification scheme then calls
%for the $n-1$ threshold classical secret sharing protocols,
%denoted by $(k,n)$ with $2\le k\le n$.
%The total correlation (\ref{tc}) in an $n$-party state
%then simply equals to the capacity of the
% $(2,n)$ threshold classical secret sharing protocol.
%It is a sum of correlations of $k$-parties with
%$k=2,3,4,\cdots,n$. The $k$-party
%correlation is defined as the difference in capacities
%between the
%$(k,n)$ and $(k+1,n)$ threshold classical secret sharing protocols,
%and the $n$-party correlation is simply
%the capacity of the $(n,n)$ threshold classical secret
%sharing protocol.

Although an efficient algorithm remains to be found to
compute the hierarchy of correlations
for a general $n$-party state, surprisingly we find
these calculations can be performed analytically
for all graph states.

An $n$-qubit graph state can be represented by a fully
connected $n$-vertex graph.
It is defined by an abelian subgroup $S_n$
of the $n$-qubit Pauli group $G_n$ with $n$ generators.
$S_n$ is called the stabilizer group
because the graph state
it defines is invariant when acted upon by its elements.
A complete set of independent elements of
$S_n$ forms the generator of the group, denoted by $\langle S_n \rangle$.
Despite the rich variety of choices
for the generator $\langle S_n \rangle$,
the number of elements in $\langle S_n \rangle$,
denoted by $|\langle S_n \rangle|$,
is definite and equals $n$ for a $n$-qubit graph state.

The total correlation $(\ref{tc})$
for an $n$-qubit graph state is then simply
equal to $|\langle S\rangle|=n$ since the one-qubit
reduced density matrix is uniformly a completely
mixed state for all qubits.
Each element in the stabilizer then
represents $1$ bit of correlation.
To compute the capacity of the
$(k,n) \;(2\le k\le n)$ threshold
classical SS in an $n$-qubit graph state,
we classify the elements in $S_n$ to the sets
$S_{k}\; (2\le k\le n)$, with $S_{k}$ composed of
 all elements in $S_n$
containing not more than $k$ single qubit Pauli matrices
distinct from identity. Clearly we have
$S_{2}\subseteq S_{3}\subseteq S_{4}
\cdots \subseteq S_{n}$.
The elements in $S_k$ can be easily shown
to represent reduced density matrices of not more than $k$-parties,
thus can be used to share secrets
among not more than $k$ parties.
In general the set $S_k$ is not a group, but its
independent elements in $S_k$ can be used
to generate a stabilizer group whose
stabilizer is denoted by $\langle S_k\rangle$.
The $k$-party correlation $C_k$ then is equal to
\begin{eqnarray}
C_T(\langle S_{k} \rangle)-C_T(\langle S_{k-1} \rangle)=|\langle
S_{k} \rangle|-|\langle S_{k-1} \rangle|. \label{eq10}
\end{eqnarray}

As an example, we list our results on the correlation structure for
all five-qubit graph states in Table \ref{table1}.
The total correlation of all five-qubit graph states is $5$ bits,
which distributes differently to among the $2$-, $3$-, $4$-,
and $5$-party correlations for different states as listed.
For instance, we note that only the first graph
state contains $1$ bit of $5$-party correlation,
while only the second graph
state has $1$ bit of $4$-party correlation, and the last graph state
has $5$ bits of $3$-party correlation. We emphasize that this
classification of correlation structure is
locally unitary invariant. Furthermore, the end result on
the correlation structure
is independent of the specific labels for the qubits.
Thus our classification scheme allows for a transparent
categorization of graph states into
local unitary equivalent classes.
The example of the 5-qubit graph state
above shows that the $2$-party (or $3$-party)
correlation alone is enough to distinguish
all four distinct classes of five-qubit graph states.
We thus state as a conjecture here that the correlation
structure for any $n$-qubit graph state
is sufficient to distinguish
different local unitary equivalent classes.
More detailed discussion on this will be given elsewhere.
\begin{table}
\begin{tabular}{|c|c|c|c|c|}
\hline & \includegraphics[width=1.5cm,height=1.5cm]{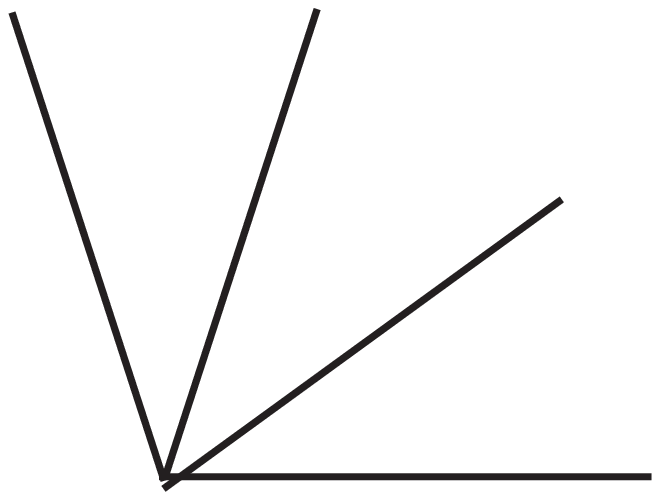}&
\includegraphics[width=1.5cm,height=1.5cm]{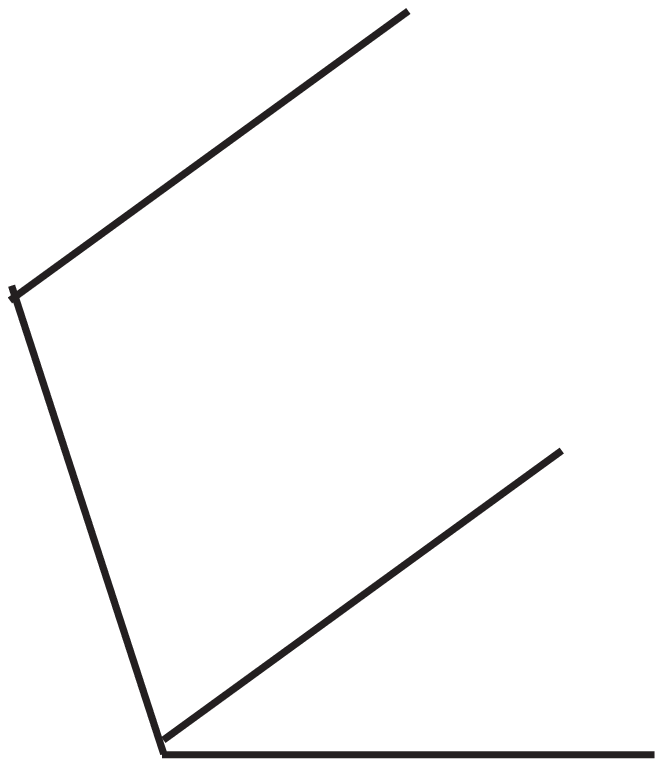}
& \includegraphics[width=1.5cm,height=1.5cm]{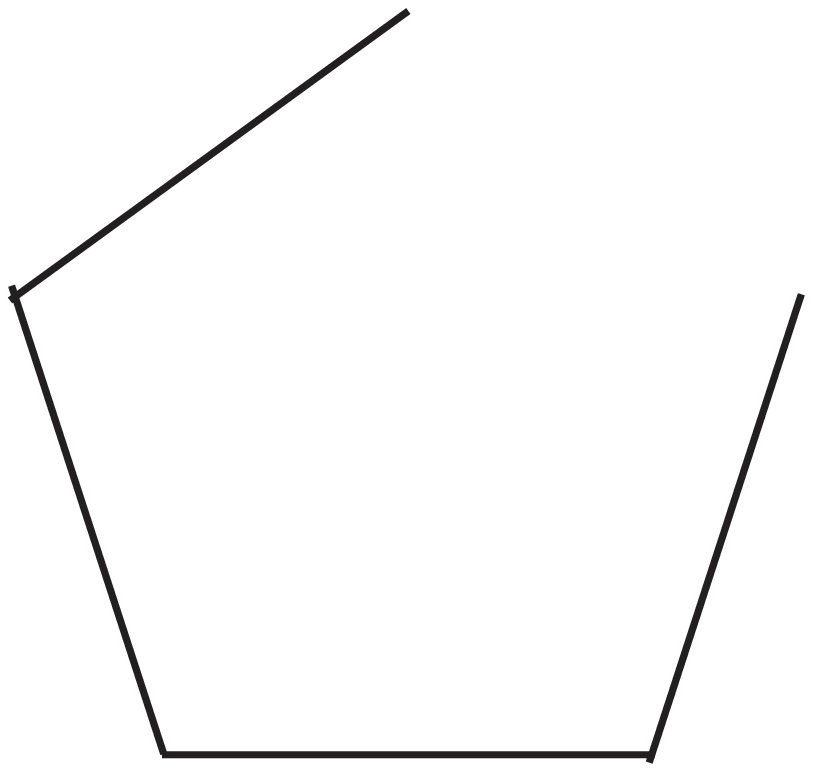}
& \includegraphics[width=1.5cm,height=1.5cm]{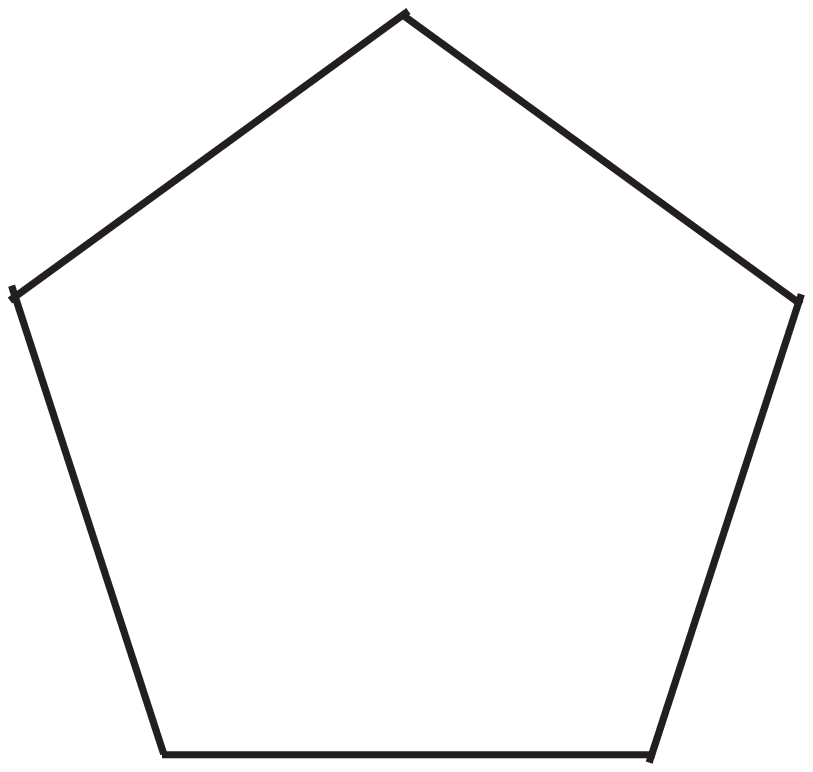}\\
\hline
 $C_2$&4&3&2&0\\
 \hline
 $C_3$&0&1&3&5\\
 \hline
 $C_4$&0&1&0&0\\
 \hline
 $C_5$&1&0&0&0\\
\hline
\end{tabular}
\caption{The correlation structure for all five-qubit graph
states.} \label{table1}
\end{table}

Before concluding, we provide further digestion of
our result by comparison with a related recent
study of multiparty quantum entanglement in an
n-qubit graph state \cite{Fattal}.
We find that their main result-$2$ \cite{Fattal}
can be obtained directly from our classification
scheme based on Eq. (\ref{eq10}), provided
the appropriate association of
the $k$-party correlation in the $k$-party
$n$-qubit graph state is taken.
Our classification scheme, on the other hand,
is more powerful and complete.
In addition, it provides a transparent
picture in terms of capacities of threshold
SS protocols.

In summary, we have proposed a scheme to characterize
the complete correlation structure in an $n$-party quantum state
based on the state's capacities for $(k,n)$ threshold
classical SS protocols ($2\le k\le n$).
The total correlation in an $n$-party state is then
found to be equal to
the capacity of classical SS in a $(2,n)$
threshold protocol, which is the same as the
sum of every single-party entropy minus the entropy for the whole
state. This total correlation is further classified into
constituents of $k$-party
($2\le k\le n$) correlations, with the $k$-party ($2\le k\le $n$-1$)
correlation being the capacity difference between the
$(k,n)$ and $(k+1,n)$ threshold protocols,
and the $n$-party correlation in an $n$-party state is defined as
the capacity of the $(n,n)$ threshold SS
protocol.  Our result allows for an
easy explanation of why the three-party mutual entropy for
a three-party state can take negative values,
thus mutual entropy cannot represent a
legitimate three-party correlation measure.
We have provided general results on the complete
correlation structure for an $n$-qubit graph state,
and give an explicit construction for the case
of five-qubit graph states.
Finally we note that the
$k$-party entanglement measure proposed by Fattal \textit{et. al.},
is simply the $k$-party correlation in a $k$-party $n$-qubit graph state.

The author (D.L.Z.) thanks B. Zeng for many useful discussions. This
work is supported by NSFC and NSF.

\end{document}